# An Evolutionary Approach to Associative Memory in Recurrent Neural Networks

**Shigetaka Fujita**
*Graduate School of Science and Technology*
*Kobe University*
*Rokkodai, Nada, Kobe 657, Japan*
e-mail: fujitan@phys02.phys.kobe-u.ac.jp

**Haruhiko Nishimura**
*Department of Information Science*
*Hyogo University of Education*
*Yashiro-cho, Kato-gun, Hyogo 673-14, Japan*
e-mail: HNISHIMU@JPNYITP.BITNET

**Abstract**
In this paper, we investigate the associative memory in recurrent neural networks, based on the model of evolving neural networks proposed by Nolfi, Miglino and Parisi. Experimentally developed network has highly asymmetric synaptic weights and dilute connections, quite different from those of the Hopfield model. Some results on the effect of learning efficiency on the evolution are also presented.

# 1 Introduction

As a model of associative memory in terms of the recurrent-type artificial neural networks, Hopfield model [1] and its variants have been successfully developed and made some important results using the statistical mechanical techniques on the analogy with spin glass [2, 3]. Their assumption of the symmetrically and fully connected weights of synapses is, however, unrealistic, since this does not arise naturally in biological neural systems. Some researches are found allowing asymmetric synaptic weights or diluting the network connections [2, 3]. They do not concern the network structures, but most of them merely investigate the effects on the storage capacities under limited conditions.

In any of above mentioned models, the physical positions where neurons exist or the distances between neurons are not considered at all. It must play an important role for the formation of biological networks in the brain. There should be the attempt to construct more natural models based on biologically and genetically plausible considerations, in which asymmetric and dilute connections are just the result. The genetic algorithm [4] can be a key ingredient toward such directions. Recently evolutionary search procedures have been combined with the artificial neural networks (so called "the evolutionary artificial neural networks") [5]. Almost all of the models deal with the feedforward multi-layered networks to raise the performance of the networks and in many of them the evolution of both weights and architectures is not integrated.

In this paper we apply a genetic evolutionary scheme to recurrent neural networks and investigate associative memory. Aiming to construct a biologically inspired neural network, we take into account the following two points in network formation:

- The correspondence of genotype and phenotype should not be direct. Only developmental rules of neurons are encoded in the genotype representation.

- The information on the physical positions of neurons and the distances between neurons should be included in the model. The resulting network is a physical object in space.

While the first point, the indirect encoding scheme, has been studied [6, 7], few attempts have been made to incorporate both of the above two points, except for the model of "evolving (growing) neural network" by Nolfi, et al. [8, 9]. So we follow their scheme treating a neural network as a creature in a physical environment.

The outline of our model is as follows. (1) Prepare the individuals which have the genotypes encoded the information for generating networks. (2) Learn the patterns according to an appropriate learning rule. (3) Impose the task for retrieving the given patterns. (4) Some individuals which make good results in the retrieval task are selected and inherit the genotypes to their offspring with mutation. (5) Repeat (2)-(4) until appropriate generation. As the generations go on, the abilities of the individuals will improve.

## 2 Model descriptions

A population consists of M individuals. Each individual has a genotype encoded the information of creating the network structure. The genotype is divided into N blocks corresponding to N neurons. Each block contains the instructions for one neuron, namely, the "physical position" of the neuron ($x$ and $y$ coordinates in the "physical space", 2 dimensional plane in this paper), the directional angle $\alpha$ for the "axonal" growth, the axonal bifurcation angle $\theta$, the axonal bifurcation length $d$, the "presynaptic" weight $b$ (characterize the firing intensity and neuron's type, excitatory or inhibitory) and the "postsynaptic" weight $a$ (amplify or damp the incoming signal) (Fig. 1). Note that these are essentially local information peculiar to each neuron. There is no direct information as to the connections between neurons. All values are randomly generated at the 0-th generation.

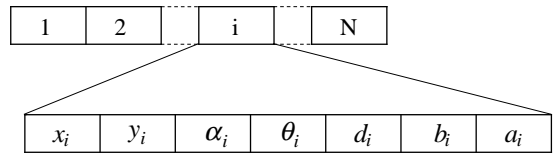

Figure 1: A Genotype consisted of N blocks and parameters specified in each block.

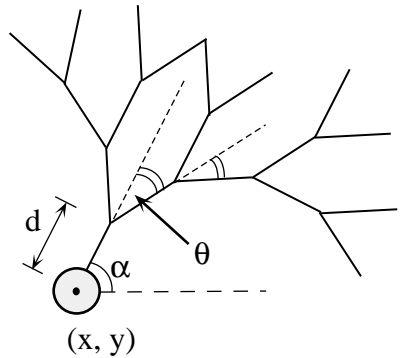

Figure 2: The growing and bifurcating process of an axon.

From each neuron in the physical space, an axon grows and bifurcates as shown in Fig. 2 and connects with other neurons. The connections of neurons are determined by whether the axon reaches the neighbourhood $r$ of other neurons or not. For convenience, we introduce the connectivity matrix $c_{ij}$ ($i, j = 1, \ldots, N; c_{ii} = 0$) as $c_{ij} = 1$ if the axon from the i-th neuron reaches the j-th neuron, and otherwise $c_{ij} = 0$. Then the *bare* synaptic weight $w_{ij}^0$ at birth, before learning, is given by $w_{ij}^0 = a_i b_j$ if $c_{ij} = 1$, and $w_{ij}^0 = 0$ if $c_{ij} = 0$. So the synaptic weights are generally asymmetric ($w_{ij}^0 \neq w_{ji}^0$).

After the bare weights are established, they are *dressed* by learning for a certain period of time. That is, p patterns $\{\xi_i^\mu = \pm 1\}(\mu = 1, \ldots, p)$ to be stored are embedded into the network by an appropriate learning rule: $w_{ij}^{New} = w_{ij}^{Old} + \delta w_{ij}$. Iterated several times in one generation (learning task), the dressed weights are obtained. In general, the synaptic increment $\delta w_{ij}$ can be modulated in each learning step. Then the "retrieval task" is imposed to the neural network (phenotype) with the dressed weights $w_{ij}$. Concretely, one of p patterns with random noise is given to the individual network as the initial states of neurons. And the state of the neuron is updated synchronously according to the equation $s_i(t+1) = f(\sum_{j=1}^{N} w_{ij} s_j(t))$, where $s_i(t)(-1 \leq s_i \leq 1)$ is the state of the i-th neuron and $t$ is the time step at retrieving. The transfer function $f(x)$ is usually taken to be $f(x) = tanh(x/2\epsilon)$. During the update proceeds, we trace the overlap of the state of the network with the target pattern: $m^\mu(t) = 1/N \sum_{i=1}^{N} \xi_i^\mu s_i(t)$. This procedure is applied to all p patterns.

The retrieval task is repeated several times with changing the noise of the set of p patterns, because the proper network should have the balanced robustness against the noise. For this reason, we take the fitness as the sum of the values of overlaps over all tasks. At the end of one generation, some individuals which have got higher fitness values are selected. Their genotypes are inherited to their offspring with random mutation (we do not use crossover for simplicity). The above process continues for appropriate generations.

## 3 Simulations and results

To carry out the experiments, we have made the following choices of parameters. A population consists of 50 ($M = 50$) individuals. The best 10 of individuals which have got higher fitness values are selected, and 5 copies of each of their genotypes are inherited to their offspring with mutation. The mutation rate is 0.004. We take a square, normalized to 1×1, as the physical space. The ranges of each parameter for a genotype are $0 \leq x, y \leq 1$, $0 \leq \alpha \leq 2\pi$, $0 \leq \theta \leq \pi/3$, $0 \leq d \leq 0.15$, $-2 \leq b \leq 2$ and $0 \leq a \leq 2$. The axon bifurcates 5 times and the neighbourhood of a neuron is fixed to $r = 0.05$. And the steepness parameter $\epsilon$ in the transfer function is set to 0.015. In learning and retrieval tasks, we use the 7×7 non-orthogonal 4 ($p = 4$) patterns as a set of target patterns (Fig. 3), then the total number of neurons becomes 49 ($N = 49$).

As the learning rule we here introduce simple Hebbian rule for the connected neurons: $\delta w_{ij} = \lambda \sum_{\mu=1}^{p} \xi_i^\mu \xi_j^\mu$ if $c_{ij} = 1$, and $\delta w_{ij} = 0$ if $c_{ij} = 0$, where $\lambda$ is the "learning efficiency" represents the intensity of *a posteriori* learning after the birth. Note that we do not take into consideration the evolution of learning rule itself. In this paper we make $\lambda$ fixed through generations. Learning and retrieval task processes in one generation are described by Fig. 4. First, 15 steps are given for learning. Secondly, for each of 4 patterns 15 steps are assigned to retrieving trial by starting with the noisy pattern state with 5% random noise. Then the overlaps are marked at the step t=5,10 and 15 and summed up. The retrieval task is repeated 4 times successively (RT1∼RT4) by changing the noise.

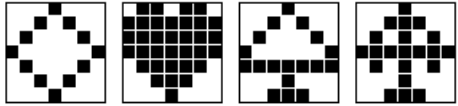

Figure 3: An example of a set of target patterns.

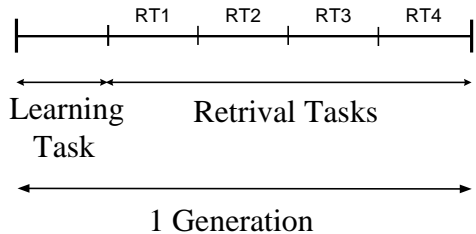

Figure 4: Flow of learning and retrieval task processes in one generation.

Bold solid lines in Fig. 5 show the transition of the fitness of the best individual until the 1000th generation at $\lambda = 0.002$ (Fig. 5(a)) and $\lambda = 0.05$ (Fig. 5(b)), respectively. The fitness value is normalized to max 100. The "bare weight" line(thin) represent the ability of the bare weights of the best individual, made by removing the learned changes from the dressed weights, namely the "native ability" for retrieving. The "pure genetic" line(dashed) represents the ability of genetic algorithm only without learning after the birth ($\lambda = 0$), namely means acquisition of the "inherited memory". Changing the learning efficiency $\lambda$, we found that higher the parameter $\lambda$ becomes, more the native ability goes down. It appears that the presence of learning helps the network to fit its environment and makes evolution easier.

Fig. 6 and Fig. 7 show the network structures of the best individual and the distributions of distance versus weight value between connected neurons generated at the 0th generation(G0) and the 1000th(G1000) in the case of $\lambda = 0.05$, respectively.

Here we cite the comparison of our numerical results with those of the conventional Hopfield model. In the following, **H** indicates the Hopfield model and **E** indicates the best individual of our evolutionary model of $\lambda = 0.05$ at the 1000th generation. The total numbers of connections are **H**:2352 and **E**:406, then the mean values of the number of connections per neuron become **H**:48 and **E**:8.29. This shows **E** develops strongly diluted connections. The values of symmetry parameter, defined by $\eta = \sum_{i,j} w_{ij} w_{ji} / \sum_{i,j} w_{ij}^2$, are given by **H**:1.0 and **E**:0.23. This means **E**

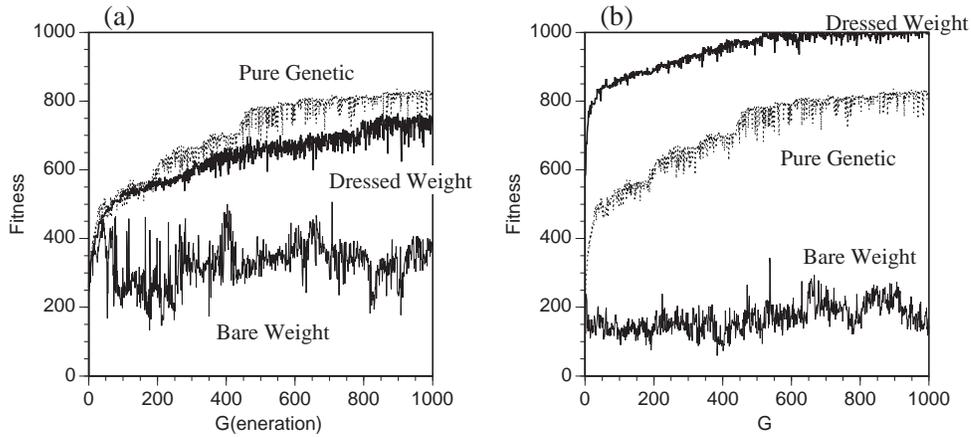

Figure 5: Evolution of fitness of the best individual in population. (a) $\lambda = 0.002$. (b) $\lambda = 0.05$.

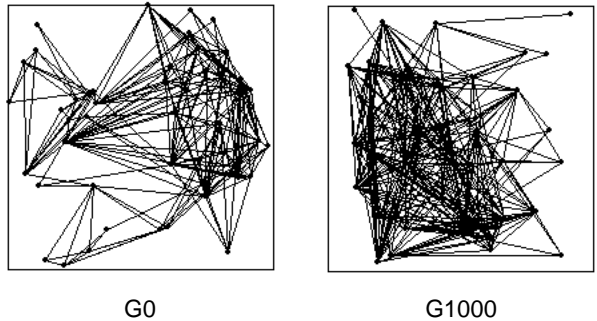

Figure 6: Network structures of the best individual at G0 and G1000 ($\lambda=0.05$).

forms highly asymmetric weights.

Finally we mention about the fitness values. **E** can reach 100 fitness value for noisy(5%) patterns, but **H** cannot reach 100 and gets 88.3 in average naively (remember that we admit non-orthogonal patterns). And **E** is almost stable for sufficient retrieving time. This suggests that our evolutionary approach may exceed the Hopfield model in associative memory, in spite of quite different structure of networks.

## 4  Summary

We have presented the results of simulations of associative memory in the recurrent-type neural networks composed of growing neurons. We found that the developed networks are far different from the Hopfield model in respect of the symmetricity and the connection number per neuron of the connection weights, but it is possible to act as a sufficient or a better associative memory than the conventional Hopfield model.

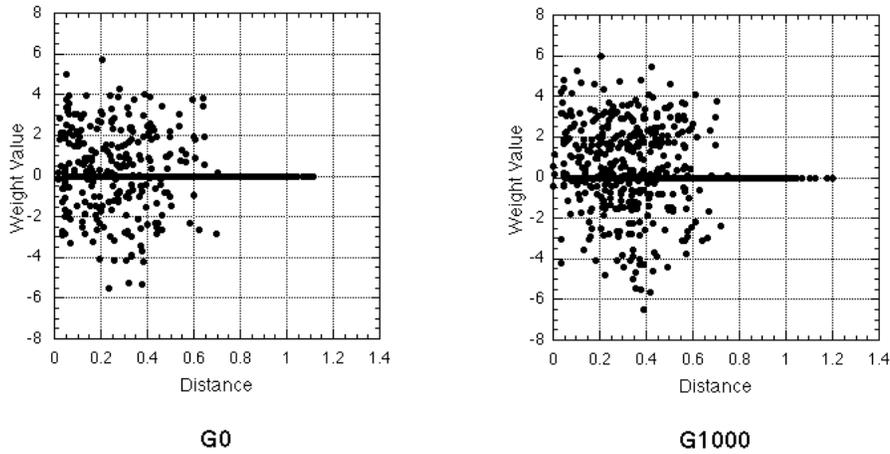

Figure 7: Dressed weight value versus distance between connected neurons of the best individual at G0 and G1000 ($\lambda=0.05$).

In this paper we kept our model simple, in terms of minimizing computational requirements and studying the feasibility of the methodology of biologically inspired associative memory models.

Further simulations are going to concentrate on the influence of increasing the sets of task-patterns. It is expected that the dynamic environment where diverse tasks are present excludes the task-specificity and makes us possible to examine the relationship between learning and evolution [10, 11].

It seems interesting as a future direction of research to incorporate epigenetic factors [12] such as the primary processes of neuronal development (adhesion, apoptosis etc.) and/or the action of nerve growth factor from target neurons.